\documentclass[A4paper,twocolumn]{article}
\usepackage{graphicx}
\usepackage{amsmath}

\begin{document}

\title{Entanglement Criterion for Coherent Subtraction and Coherent Addition
Bipartite Continuous variable States }
\author{Li-zhen Jiang$^1$, Xiao-yu Chen$^1$, Tian-yu Ye$^1$, Fang-yu Hong $^2$ \\
{\small {$^1$ College of Information and Electronic Engineering, Zhejiang
Gongshang University, Hangzhou, Zhejiang 310018, China }}\\
{\small {$^2$ Department of Physics, Zhejiang Sci-Tech University, Hangzhou,
Zhejiang 310018, China }}}
\date{}
\maketitle

\begin{abstract}
Photon subtraction and addition are experimental means of generating
non-Gaussian states from Gaussian states. Coherent subtraction or addition
is a combination of photon subtractions or additions. The resultant states
are quite general non-Gaussian states. The states can be photon number
entangled states with arbitrary coefficients. We derive the entanglement
conditions for several classes of coherent subtraction or coherent addition
bipartite continuous variable states. One of the entanglement conditions is
necessary and sufficient.

PACS number(s): 03.67.Mn; 03.65.Ud\newline
\end{abstract}

\section{Introduction}

Quantum continuous variable entanglement is essential for quantum
information tasks realized with quantum optical means. A great deal of
contributions have been devoted to the entanglement criteria of Gaussian
states \cite{Duan} \cite{Simon} \cite{Wang} \cite{Giedke} \cite{Werner}
which are continuous variable states that can be determined only by their
first and second moments. Recently, non-Gaussian states attract much
attentions for its good performance in quantum teleportation \cite{Opatrny},
dense coding \cite{Kitagawa}, quantum computation \cite{Bartlett} and
nonlocality test \cite{Nha04}. Several entanglement criterion have also been
proposed for non-Gaussian states \cite{SV} \cite{Hillery} \cite{Agarwal}
\cite{Chen07} \cite{Nha}. However, types of non-Gaussian states that have
been studied are sporadic and limited by experimental conditions and
theoretical interest until now. We will introduce the general ideas of
coherent subtraction and addition to describe non-Gaussian states and study
their entanglement conditions.

To generate non-Gaussian state from Gaussian state, one of the means is to
subtract (or add) photon. The photon subtraction (addition) state is $%
a_1^ma_2^{m^{\prime }}\rho _Ga_1^{\dagger m}a_2^{\dagger m^{\prime }}$ ($%
a_1^{\dagger m}a_2^{\dagger m^{\prime }}\rho _Ga_1^ma_2^{m^{\prime }}$),
where $\rho _G$ is a Gaussian state (hereafter referred as Gaussian kernel)
, $a_1$ and $a_2$ are the annihilation operators of the two modes, $%
a_1^{\dagger }$ and $a_2^{\dagger }$ the creation operators. The replacement
of $a_i$ with the superposition $ta_i+ra_i^{\dagger }$ has been proposed for
quantum state engineering \cite{Lee}. The most generic definition of
bipartite photon coherent subtraction and addition state is $\rho
_{SA}=C_{SA}\rho _GC_{SA}^{\dagger },$ where $C_{SA}=%
\sum_{m_1,m_2,m_3,m_4=0}^\infty A_{m_1m_2m_3m_4}a_1^{\dagger
m_1}a_2^{\dagger m_2}a_1^{m_3}a_2^{m_4},$ with complex coefficients $%
A_{m_1m_2m_3m_4}.$ This is so generic a definition that all two-mode
continuous variable states can be expressed in such form. The photon
coherent subtraction state is $\rho _S=C_S\rho _GC_S^{\dagger },$ with $%
C_S=\sum_{m_1m_2}A_{m_1m_2}a_1^{m_1}a_2^{m_2}.$ The photon coherent addition
state is $\rho _A=C_A\rho _GC_A^{\dagger },$ with $C_A=%
\sum_{m_1m_2}A_{m_1m_2}a_1^{\dagger m_1}a_2^{\dagger m_2}.$ More
specifically, if the two modes are symmetry, the symmetric coherent
subtraction and addition states are
\begin{eqnarray*}
\rho _{SS} &=&\sum_{m=0}^\infty A_ma_1^ma_2^m\rho _G^{II}\sum_{n=0}^\infty
A_n^{*}a_1^{\dagger n}a_2^{\dagger n}, \\
\rho _{AS} &=&\sum_{m=0}^\infty A_ma_1^{\dagger m}a_2^{\dagger m}\rho
_G^{II}\sum_{n=0}^\infty A_n^{*}a_1^na_2^n,
\end{eqnarray*}
respectively. The symmetric coherent subtraction or coherent addition state
set is still large enough to contain many interesting non-Gaussian states.
In this paper, we will study the entanglement of several symmetric coherent
subtraction or addition states. Section 2 is to introduce the realignment
entanglement criterion for non-Gaussian states, with a new form of the
criterion in coherent state representation. Section 3 is devoted to the
entanglement conditions of coherent subtraction state when the Gaussian
kernel is a two-mode squeezed thermal state (TMST), when the Gaussian kernel
is in its standard form. In section 4, we investigate the entanglement
conditions of coherent addition states such as photon number entangled
states (PNES) evolved in thermal noise and amplitude damping environment,
coherent addition of TMST. Conclusions are drawn in section 5.

\section{Realignment entanglement criterion in coherent state representation}

The realignment entanglement criterion has been studied for continuous
variable system recently \cite{Chen}. The derivation is based on Fock space
representation. We here give the criterion based on coherent state
representation. A bipartite continuous variable state $\rho $ is completely
specified by its characteristic function $\chi (\mu )=Tr[\rho \mathcal{D}%
(\mu )],$ where $\mathcal{D}(\mu )=\exp (\mu a^{\dagger }-\mu ^{*}a)$ is the
displacement operator, and $a=(a_1,a_2)^T,$ $\mu =(\mu _1,\mu _2)$ for $%
1\times 1$ two-mode state$.$ A Gaussian state is completely determined by
its first and second moments. The characteristic function of a Gaussian
state $\rho _G$ with nullified first moments (the first moments are not
relevant to the entanglement and can be removed by displacement) is $\chi
_G(\mu )=\exp [-\frac 12(\mu ,\mu ^{*})\gamma (\mu ,\mu ^{*})^T],$ where $%
\gamma $ is the $4\times 4$ complex covariance matrix of $\rho _G.$ Denote $%
\alpha =(\alpha _1,\alpha _2),$ and denote the two mode coherent state as $%
\left| \alpha \right\rangle =\left| \alpha _1,\alpha _2\right\rangle $. In
coherent state representation, we have $\left\langle \alpha \right| \rho
_G\left| \beta \right\rangle =\int \left[ \prod_{i=1}^2\frac{d^2\mu _i}\pi
\right] \chi _G(\mu )\left\langle \alpha \right| \mathcal{D}(-\mu )\left|
\beta \right\rangle =$ $\frac 1{\sqrt{\det (\gamma ^{\prime })}}\exp [-\frac{%
\left| \alpha \right| ^2+\left| \beta \right| ^2}2+\frac 12(\alpha ^{*},$ $%
\beta )$ $\Gamma $ $(\alpha ^{*},$ $\beta )^T].$ Where $\Gamma =\sigma
_1\otimes I_2+\sigma _3\otimes I_2\gamma ^{\prime -1}\sigma _3\otimes I_2,$ $%
\gamma ^{\prime }=\gamma +\frac 12\sigma _1\otimes I_2.$ Here $\sigma _i\ $%
are the Pauli matrices, $I_2$ is the $2\times 2$ identity matrix. The
realigned density matrix $\rho _{GR}$ is the interchange of second and third
subscripts of $\rho _G,$ namely, $\left\langle \alpha _1\alpha _2\right|
\rho _{GR}\left| \beta _1\beta _2\right\rangle =\left\langle \alpha _1\beta
_1\right| \rho _G\left| \alpha _2\beta _2\right\rangle .$ Thus
\[
\left\langle \alpha \right| \rho _{GR}\left| \beta \right\rangle =\frac 1{%
\sqrt{\det \gamma ^{\prime }}}\exp [-\frac{\left| \alpha \right| ^2+\left|
\beta \right| ^2}2+\frac 12(\alpha ^{*},\beta )\Gamma _R(\alpha ^{*},\beta
)^T]
\]
where $\Gamma _R=Z\Gamma Z,$ with $Z_{11}=Z_{44}=Z_{23}=Z_{32}=1$ and all
the other entries of the $4\times 4$ matrix $Z$ are zeros. The realignment
criterion of entanglement is $Tr\sqrt{\rho _R\rho _R^{\dagger }}>1$, from
which we can derive a condition of entanglement$Tr\rho _R>1.$ The later is
easy for use but may detect less number of entangled states as the former.
Here $\rho _R$ is the realignment of $\rho .$ For a Gaussian state, we have
the realignment entanglement sufficient condition
\[
Tr\rho _{GR}=\sqrt{\frac{\det \gamma _R^{\prime }}{\det \gamma ^{\prime }}}%
>1,
\]
where $\gamma _R^{\prime }=\sigma _3\otimes I_2(\Gamma _R-\sigma _1\otimes
I_2)^{-1}\sigma _3\otimes I_2.$

The non-Gaussian state produced by coherent subtraction and addition from
Gaussian state can be written as derivative of functional of Gaussian state.
The method to derive the entanglement criterion of non-Gaussian states can
be found in Ref.\cite{Chen}.

\section{Coherent subtraction}

A $1\times 1$ bipartite Gaussian state can be transformed to its first
standard form (denoted as $\rho _G^I$) by local operations. Further, it is
always possible to transform $\rho _G^I$ to its second standard form $\rho
_G^{II}\ $by proper local squeezing operations\cite{Duan}. To prove the
necessary condition of entanglement for a Gaussian state, it is shown that a
separable Gaussian state $\rho _G^{II}$ takes the form of
\begin{equation}
\rho _G^{II}=\int P(\alpha _1,\alpha _2)\left| \alpha _1,\alpha
_2\right\rangle \left\langle \alpha _1,\alpha _2\right| d^2\alpha
_1d^2\alpha _2,  \label{wee1}
\end{equation}
where $P(\alpha _1,\alpha _2)$ is a Gaussian function hence positive
definite, $\left| \alpha _1,\alpha _2\right\rangle $ is the product of
coherent states.

Photon subtraction from Gaussian state $\rho _G^{II}$ will lead to a
non-Gaussian state $a_1a_2\rho _G^{II}a_1^{\dagger }a_2^{\dagger }.$ More
generic photon subtraction is the coherent subtraction, or the superposition
of all possible subtraction operations, namely, $\sum_{m,m^{\prime
}=0}^\infty A_{mm^{\prime }}a_1^ma_2^{m^{\prime }}$. The non-Gaussian state
produced from Gaussian kernel $\rho _G^{II}$ is
\begin{equation}
\rho _S=\sum_{m,m^{\prime }=0}^\infty A_{mm^{\prime }}a_1^ma_2^{m^{\prime
}}\rho _G^{II}\sum_{n,n^{\prime }}A_{nn^{\prime }}^{*}a_1^{\dagger
n}a_2^{\dagger n^{\prime }}.  \label{wee2}
\end{equation}
Then state $\rho _S$ is separable when $\rho _G^{II}$ is separable, since we
have
\begin{eqnarray}
\rho _S &=&\int P(\alpha _1,\alpha _2)\left| \sum_{m,m^{\prime }=0}^\infty
A_{mm^{\prime }}\alpha _1^m\alpha _2^{m^{\prime }}\right| ^2  \nonumber \\
&&\times \left| \alpha _1,\alpha _2\right\rangle \left\langle \alpha
_1,\alpha _2\right| d^2\alpha _1d^2\alpha _2,  \label{wee3}
\end{eqnarray}
which is explicitly separable. Hence a non-Gaussian state prepared by
coherent subtraction from second standard form Gaussian state is separable
if the kernel Gaussian state is separable. A Gaussian state with first
standard form $\rho _G^I$ differs from its second standard form by local
squeezing, $\rho _G^I=S_1(r_1)S_2(r_2)\rho _G^{II}S_1^{\dagger
}(r_1)S_2^{\dagger }(r_2)$ with some properly chosen squeezing parameters $%
r_1$ and $r_2,$ and $S_1,S_2$ are the single-mode squeezing operators. Keep
in mind that the explicit separable decomposition (\ref{wee1}) is only for
the Gaussian state with second standard form. So $\rho _G^I$ may not be
diagonal in coherent state representation even when it is separable.

\subsection{Coherent subtraction of two mode squeezed thermal state}

A TMST is $\rho _{st}=S(r)\rho _{th}S^{\dagger }(r)$, where $S(r)=\exp
[r(a_1^{\dagger }a_2^{\dagger }-a_1a_2)]$ is the two mode squeezing
operator, $\rho _{th}=\rho _{th,1}\otimes \rho _{th,2}$ is the thermal state
with $\rho _{th,i}=(1-v_i)\sum_nv_i^n\left| n\right\rangle \left\langle
n\right| .$ The average photon number of state $\rho _{th,i}$ is $%
N_i=v_i/(1-v_i).$ Consider the case of symmetric modes, we have $%
N_1=N_2\equiv N.$ The characteristic function of $\rho _{st}$ is $\chi
_{st}(\mu )=\exp [-\frac 12(\mu ,\mu ^{*})\gamma (\mu ,\mu ^{*})^T],$ where
the complex covariance matrix $\gamma =b_0\sigma _1\otimes I_2+c_1I_2\otimes
\sigma _1,$ with $b_0=(N+\frac 12)\cosh 2r,$ $c_1=-(N+\frac 12)\sinh 2r.$
The characteristic function can also be written as $\chi (\mu _1,\mu
_2)==\exp [-\frac 12(\mu _1^I,\mu _1^R,\mu _2^I,\mu _2^R)M(\mu _1^I,\mu
_1^R,\mu _2^I,\mu _2^R)^T],$ where $\mu _i^I$ and $\mu _i^R$ are imaginary
and real parts of $\mu _i.$ The real symmetric correlation matrix is $M=$ $%
2b_0I_2\otimes I_2-2c_1\sigma _1\otimes \sigma _3.$ Notice that the first
standard form $M^I$ and the second standard form $M^{II}$ are equal to $M$%
\cite{Duan} for symmetric TMST$.$ We have $\rho _{st}^I=\rho _{st}^{II}=$ $%
\rho _{st}$ for symmetric TMST. The necessary and sufficient condition of
the separability of the symmetric TMST is
\begin{equation}
2b_0+2c_1\geq 1.  \label{wee4}
\end{equation}
Lets consider coherent subtraction state $\rho _{Sst}=\sum_{m,m^{\prime
}}A_{mm^{\prime }}a_1^ma_2^{m^{\prime }}\rho _{st}\sum_{n,n^{\prime
}}A_{nn^{\prime }}^{*}a_1^{\dagger n}a_2^{\dagger n^{\prime }}.$ If (\ref
{wee4}) is fulfilled, $\rho _{st}$ is separable, then $\rho _{Sst}$ is
separable since it can be written in the form of Eq.(3).

The inverse problem is that if $\rho _{Sst}$ is entangled or not when $\rho
_{st}$ is entangled. We consider the case of symmetric coherent subtraction
of photon, the symmetric non-Gaussian state is $\rho
_{SSst}=\sum_mA_ma_1^ma_2^m\rho _{st}\sum_nA_n^{*}a_1^{\dagger
n}a_2^{\dagger n}.$ The realignment entanglement criterion for unnormalized $%
\rho _{SSst}$ is \cite{Chen}
\begin{equation}
\sqrt{\tau }\mathcal{O}P^R>\mathcal{O}P.  \label{wee5}
\end{equation}
where $\tau =\frac 1{4(b_0+c_1)^2}$ ,$\mathcal{O}=\sum_{m,n}A_mA_n^{*}\left.
\frac{\partial ^{2m+2n}}{\partial \xi _1^m\partial \xi _2^m\partial \eta
_1^n\partial \eta _2^n}\right| _{\xi _i=\eta _i=0},$ $P^R=\exp [L(\eta _1\xi
_1+\eta _2\xi _2)+K(\eta _1\eta _2+\xi _1\xi _2)],$ $P=\exp [(b_0-\frac
12)(\eta _1\xi _1+\eta _2\xi _2)-c_1(\eta _1\eta _2+\xi _1\xi _2)],$ with $%
K=\frac 12[(b_0+c_1)\tau +(b_0-c_1)-1],$ $L=\frac 12[(b_0-c_1)-(b_0+c_1)\tau
].$ Then The entanglement condition is $2b_0+2c_1<1,$ so we keep $b_0-c_1$
invariant. Denote $\epsilon =1-2b_0-2c_0,$ then $\epsilon $ is a small
positive quantity for entangled symmetric TMST near the separable boundary.
Expanding all the quantities in (\cite{wee5}) to the first order of $%
\epsilon ,$ we have $b_0+c_0=\frac 12-\epsilon ,$ $\tau \approx 1+4\epsilon
, $ $K=z+\frac \epsilon 2,$ $L=z-\frac \epsilon 2,$ $b_0-\frac 12=z-\frac
\epsilon 2,c_1=-z-\frac \epsilon 2,$ where $z=\frac 12(b_0-c_1-\frac 12).$
Notice that up to the first order of $\epsilon $, we have $\mathcal{O}P^R=%
\mathcal{O}P$, while $\sqrt{\tau }>1.$ Hence (\ref{wee5}) is fulfilled even
for infinitive small $\epsilon .$ Thus $\rho _{SSst}$ is entangled if $\rho
_{st}$ is entangled. We have proven that for any symmetric coherent
subtraction state $\rho _{SSst}=\sum_mA_ma_1^ma_2^m\rho
_{st}\sum_nA_n^{*}a_1^{\dagger n}a_2^{\dagger n}$ produced from symmetric
TMST, the necessary and sufficient of separability is (\ref{wee4}).

\subsection{Symmetric coherent subtraction of standard form Gaussian state}

The first standard form Gaussian state $\rho _G^I$ is completely
characterized by its complex covariance matrix $\gamma =b_0\sigma _1\otimes
I_2+c_1I_2\otimes \sigma _1+c_2\sigma _1\otimes \sigma _1,$where $%
b_0=Tr[\rho _G(a_1^{\dagger }a_1+\frac 12)]$ $=Tr[\rho _G(a_2^{\dagger
}a_2+\frac 12)],$ $c_1=Tr[\rho _Ga_1a_2],$ $c_2=Tr[\rho _Ga_1^{\dagger
}a_2]. $ The real symmetric correlation matrix is $M^I=$ $2b_0I_2\otimes
I_2-2c_1\sigma _1\otimes \sigma _3+2c_2\sigma _1\otimes I_2.$ The necessary
and sufficient condition of the entanglement of $\rho _G^I$ is $\tau =\frac
1{4(b_0+c_1)^2-c_2^2}>1.$ For symmetric coherent subtraction state $\rho
_{SS}=\sum_mA_ma_1^ma_2^m\rho _G\sum_nA_n^{*}a_1^{\dagger n}a_2^{\dagger n},$%
we consider if it is separable at the $\tau =1$ and $c_2\neq 0,$ the
boundary of separability of its Gaussian kernel. The entanglement criterion
is
\begin{equation}
\mathcal{O}P^R>\mathcal{O}P,  \label{wee6}
\end{equation}
where $P^R=\exp [b(\eta _1\eta _2+\xi _1\xi _2)-c_1(\eta _1\xi _1+\eta _2\xi
_2)+c_2(\eta _1\xi _2+\eta _2\xi _1)],$ $P=\exp [-c_1(\eta _1\eta _2+\xi
_1\xi _2)+b(\eta _1\xi _1+\eta _2\xi _2)+c_2(\eta _1\eta _2+\xi _1\xi _2)],$
with $b=$ $b_0-\frac 12.$ Then
\begin{eqnarray*}
\mathcal{O}P^R &=&\sum_{m,n}m!n!A_mA_n^{*}\sum_{k=0}^{\min (m,n)}\left(
\begin{array}{l}
m \\
k
\end{array}
\right) \left(
\begin{array}{l}
n \\
k
\end{array}
\right) c_2^{2k} \\
&&\sum_{l=0}^{\min (m,n)-k}\left(
\begin{array}{l}
m-k \\
l
\end{array}
\right) \left(
\begin{array}{l}
n-k \\
l
\end{array}
\right) b^{n+m-2k-2l}c_1^{2l}.
\end{eqnarray*}
The same expression can be found for $\mathcal{O}P$ with the interchange of $%
b$ and $-c_1.$ Consider the function $f_{m^{\prime }n^{\prime
}}(x)=\sum_{l=0}^{\min (m^{\prime },n^{\prime })}\left(
\begin{array}{l}
m^{\prime } \\
l
\end{array}
\right) \left(
\begin{array}{l}
n^{\prime } \\
l
\end{array}
\right) (x^{2l}-x^{m^{\prime }+n^{\prime }-2l}).$
Notice that $f_{m^{\prime }m^{\prime }}(x)\equiv 0,$ we only need to
consider $m^{\prime }\neq n^{\prime }.$ For odd $m^{\prime }+n^{\prime },$
one of the roots for equation $f_{m^{\prime }n^{\prime }}(x)=0$ is $x=1;$
For even $m^{\prime }+n^{\prime },$ $x=\pm 1$ are the roots of equation $%
f_{m^{\prime }n^{\prime }}(x)=0.$ We have verified numerically that there
are no other real roots for all $m^{\prime },n^{\prime }\leq 200.$ Since $%
f_{m^{\prime }n^{\prime }}(0)=1,$ we have $f_{m^{\prime }n^{\prime }}(x)>0$
for all $x\in (-1,1).$ From $\tau =1,$ we have $b+c_1=\sqrt{c_2^2+\frac 14}%
-\frac 12>0$, thus $-\frac{c_1}b\in (0,1)$ as $c_1$ is assumed to be
negative (for positive $c_1$, the realignment criterion $Tr\rho _R>1$ should
be modified \cite{Chen}), then $f_{m^{\prime }n^{\prime }}(-\frac{c_1}b)>0.$
At $\tau =1,$ the realignment entanglement criterion $\mathcal{O}P^R-%
\mathcal{O}P>0$ can be rewritten as
\begin{eqnarray}
\sum_{m>n}B_{mn}\sum_{k=0}^{\min (m,n)}\left(
\begin{array}{l}
m \\
k
\end{array}
\right) \left(
\begin{array}{l}
n \\
k
\end{array}
\right)  \nonumber \\
\times c_2^{2k}b^{n+m-2k}f_{m-k,n-k}(-\frac{c_1}b) >0,  \label{wee7}
\end{eqnarray}
where
\begin{equation}
B_{mn}=m!n!(A_mA_n^{*}+A_nA_m^{*}).  \label{wee8}
\end{equation}
If $B_{mn}\geq 0$ for all $m,n$ and $B_{mn}>0$ at least for one pair of $m,n$
$(m\neq n),$ the coherent subtracted state $\rho _{SS}.$ One of the case is
that the real and/or imaginary parts of the coefficients $A_m$ are non
negative and at least two of $A_m$ have positive real and/or imaginary
parts. Coherent subtraction means at least two of the coefficients $A_m$ are
non zeros. We conclude that if the coefficients have positive real and/or
imaginary parts, the symmetric non-Gaussian state produced from the coherent
subtraction of an edge separable first standard form of Gaussian state is
entangled.

\section{Coherent addition}

Unlike the coherent subtraction state, the sufficient condition of
separability for a coherent addition state is difficult to be found even
when the Gaussian kernel $\rho _G$ is in its second standard form. We will
consider the sufficient condition of entanglement for symmetric coherent
addition state $\rho _{SA}=\sum_{m=0}^\infty A_ma_1^{\dagger m}a_2^{\dagger
m}\rho _G\sum_{n=0}^\infty A_n^{*}a_1^na_2^n$ in the following. An initially
prepared non-Gaussian entangled state will become separable after it
interact with the thermal noise and amplitude damping environment. The
evolution of the state could be specified by the evolution of its
characteristic function \cite{Chen06}. The realignment entanglement
criterion for time evolution state is \cite{Chen}

\begin{equation}
\sqrt{\frac{\det (\gamma _{Rt}^{\prime })}{\det (\gamma _t^{\prime })}}%
\mathcal{O}\exp [-\frac 12\mathbf{v}\Omega \mathbf{v}^T]>\mathcal{O}\exp
[-\frac 12\mathbf{v}\gamma ^{\prime }\mathbf{v}^T].  \label{wee18}
\end{equation}
where $\mathbf{v=}(\varepsilon _1,\varepsilon _2,-\zeta _1,-\zeta _2),%
\mathcal{O}=\sum_{m,n}A_mA_n^{*}\mathcal{O}_{mn},$with $\mathcal{O}%
_{mn}=\left. \frac{\partial ^{2m+2n}}{\partial \varepsilon _1^m\partial
\varepsilon _2^m\partial \zeta _1^n\partial \zeta _2^n}\right| _{\varepsilon
_i=\zeta _i=0}$ $,$ $\Omega =\gamma ^{\prime }+e^{-t}\gamma ^{\prime
}(-\gamma _t^{\prime -1}+\gamma _t^{\prime -1}Z^{\prime }\gamma
_{Rt}^{\prime }Z^{\prime }\gamma _t^{\prime -1})\gamma ^{\prime }$ with $%
\gamma _t^{\prime }=e^{-t}\gamma ^{\prime }+(\widetilde{n}%
+1)(1-e^{-t})\sigma _1\otimes I_2,$ $\gamma _{Rt}^{\prime }=(Z^{\prime
}\gamma _t^{\prime -1}Z^{\prime }+I_2\otimes \sigma _1+\sigma _1\otimes
I_2)^{-1}$ and $Z^{\prime }=(\sigma _3\otimes I_2)Z(\sigma _3\otimes I_2).$
Here $\widetilde{n}$ is the thermal noise, the amplitude damping.coefficient
is merged in the unit of time $t.$

\subsection{Photon number entangled state}

A special case of symmetric coherent addition state is the PNES when the
Gaussian kernel $\rho _G$ is the vacuum state. We have $\rho _{PNES}=\left|
\psi \right\rangle \left\langle \psi \right| $ where $\left| \psi
\right\rangle =\sum_{m=0}^\infty m!A_m\left| mm\right\rangle .$ For vacuum
Gaussian kernel, $\gamma ^{\prime }=\sigma _1\otimes I_2,$ $\gamma
_t^{\prime }=n_t\sigma _1\otimes I_2,$ $\gamma _{Rt}^{\prime }=\frac{n_t}{%
2n_t-1}[(1-n_t)I_2\otimes \sigma _1+n_t\sigma _1\otimes I_2],$ where $n_t=%
\widetilde{n}(1-e^{-t})+1.$ Then $\Omega =-zI_2\otimes \sigma _1+(1-z)\sigma
_1\otimes I_2$, with $z=e^{-t}/(2n_t-1)=e^{-t}/[2\widetilde{n}(1-e^{-t})+1].$
Denote $F=-\frac 12\mathbf{v}\Omega \mathbf{v}^T=z(\varepsilon _1\varepsilon
_2+\zeta _1\zeta _2)+(1-z)(\varepsilon _1\zeta _1+\varepsilon _2\zeta _2),$
then
\begin{eqnarray*}
\mathcal{O}_{mn}e^F &=&\frac 1{(m+n)!}\mathcal{O}_{mn}F^{m+n} \\
&=&m!n!\sum_{k=0}^{\min (m,n)}\left(
\begin{array}{l}
m \\
k
\end{array}
\right) \left(
\begin{array}{l}
n \\
k
\end{array}
\right) (1-z)^{2k}z^{n+m-2k}.
\end{eqnarray*}
We also get $\mathcal{O}_{mn}\exp [-\frac 12\mathbf{v}\gamma ^{\prime }%
\mathbf{v}^T]=m!n!\delta _{mn},$ $\sqrt{\frac{\det (\gamma _{Rt}^{\prime })}{%
\det (\gamma _t^{\prime })}}=\frac 1{2n_t-1}.$ The realignment entanglement
criterion for time evolution PNES is
\begin{equation}
\sum_{m,n}\frac{C_mC_n^{*}}{2n_t-1}\sum_{k=0}^{\min (m,n)}\left(
\begin{array}{l}
m \\
k
\end{array}
\right) \left(
\begin{array}{l}
n \\
k
\end{array}
\right) (1-z)^{2k}z^{n+m-2k}>1.  \label{wee19}
\end{equation}
where $C_m=\frac{m!A_m}{\sqrt{\sum_n\left| n!A_n\right| ^2}}.$

For a two mode squeezed thermal state evolving in amplitude damping and
thermal noise channel, the initial state is $\left| \psi \right\rangle =%
\sqrt{1-\lambda ^2}\sum_{m=0}^\infty \lambda ^m\left| mm\right\rangle ,$
thus $C_m=\sqrt{1-\lambda ^2}\lambda ^m.$ By interchanging the order of
summations in (\ref{wee19}), the entanglement criterion of the final state
reduces to
\begin{equation}
\frac{1+\lambda }{(2n_t-1)(1+\lambda -2\lambda z)}>1.  \label{wee20}
\end{equation}
Condition (\ref{wee20}) can also be derived from the well know entanglement
condition of Gaussian state, for the final state is Gaussian. This example
shows that the choice of Gaussian kernel may not be relevant for the
entanglement criterion.

For a two mode coherently correlated state (TMC) with Poisson coefficients
evolves in thermal noise and amplitude damping channel, the unnormalized
initial state is $\left| \psi \right\rangle =\sum_{m=0}^\infty \frac{\lambda
^m}{m!}\left| mm\right\rangle ,$ then $C_m=\frac{\lambda ^m}{g(\lambda )m!},$
where $g(x)=\sum_m(\frac{x^m}{m!})^2.$ The function $g(x)$ converges fast
than exponential function. The entanglement criterion (\ref{wee19}) leads to
\begin{equation}
e^{2\lambda z}g(\lambda (1-z))>(2n_t-1)g(\lambda ).  \label{wee21}
\end{equation}
The critical separable time is shown in Fig.1(a) for different noise $%
\widetilde{n}.$ Comparing with the direct numerical calculation \cite
{Allegra}, we confirm that the numerical calculation of the realignment
entanglement criterion of TMC is valid.

Another example is the evolution of photonic Bell-state $\left| \Psi
_B\right\rangle =c_0\left| 00\right\rangle +c_1\left| 11\right\rangle \ $in
thermal noise and amplitude damping environment. The entanglement criterion (%
\ref{wee19}) leads to
\begin{equation}
\left| c_0\right| ^2+2Re(c_0c_1^{*})z+\left| c_1\right| ^2(1-2z+2z^2)>2n_t-1.
\label{wee22}
\end{equation}
The critical separable time is shown in Fig.1(b), where the Simon's
criterion is also shown for comparison \cite{Simon}. The similar curve has
been obtained by Nha's criterion\cite{Nha}.

\begin{figure}[tbp]
\includegraphics[ trim=0.000000in 0.000000in -0.138042in 0.000000in,
height=1.5in, width=3.6in ]{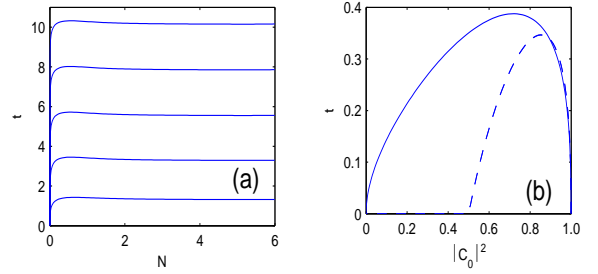}
\caption{(a)The critical damping time of separability for TMC, $N$ is the
average energy of input state, it is a little bit less than $\lambda $. From
up to bottom, the thermal noise $\widetilde{n}$=$%
10^{(}-5),10^{(}-4),10^{(}-3),10^{(}-2),10^{(}-1)$,respectively. (b)The
critical damping time of separability for photonic Bell-state. The solid
line is the realignment criterion, the dash line is the Simon's criterion.}
\end{figure}

\subsection{Coherent addition of two mode squeezed thermal state}

The coherent addition of symmetric TMST is $\rho _{ASst}=\sum_{m=0}^\infty
A_ma_1^{\dagger m}a_2^{\dagger m}\rho _{st}\sum_{n=0}^\infty
A_n^{*}a_1^na_2^n.$ The realignment entanglement criterion is still (\ref
{wee18}), where $\mathcal{O}=\sum_{m,n}A_mA_n^{*}\left. \frac{\partial
^{2m+2n}}{\partial \varepsilon _1^m\partial \varepsilon _2^m\partial \zeta
_1^n\partial \zeta _2^n}\right| _{\varepsilon _i=\zeta _i=0}$ $\Omega =\frac
12[(b_0-c_1)-(b_0+c_1)\tau ]\sigma _1\otimes I_2-\frac 12[(b_0+c_1)\tau
+(b_0-c_1)+1]I_2\otimes \sigma _1,$ $\gamma ^{\prime }=(b_0+\frac 12)\sigma
_1\otimes I_2+c_1I_2\otimes \sigma _1,$ with $\tau =\frac 1{4(b_0+c_1)^2}$ .
We consider the case that the kernel is a symmetric TMST of critical
separable, that is $\tau =1.$ Then $\Omega =-c_1\sigma _1\otimes
I_2-(b_0+\frac 12)I_2\otimes \sigma _1.$ The entanglement criterion (\ref
{wee18}) then reads
\[
\sum_{m>n}B_{mn}(b_0+\frac 12)^{n+m}f_{m,n}(-\frac{c_1}{b_0+\frac 12})>0,
\]
where $B_{mn}$ is defined as in (\ref{wee8}). Notice that when $\tau =1,$ we
have $-\frac{c_1}{b_0+\frac 12}=\frac{-c_1}{-c_1+1}\in (0,1),$ here $c_1<0$
is assumed as before. Hence $f_{m,n}(-\frac{c_1}{b_0+\frac 12})>0$ as we
have argued. So if all $B_{mn}$ are nonnegative (at least one of them is
nonzero thus it is positive), the state should be entangled. One of the
cases is that if only two of the coefficients are nonzero, say $A_m$ and $%
A_n $, the entanglement condition will be
\[
A_mA_n^{*}+A_nA_m^{*}>0.
\]
Denote $A_m=\left| A_m\right| e^{i\varphi _m},$ the entanglement condition
reduces to
\[
\cos (\varphi _m-\varphi _n)>0.
\]
for any $m\neq n.$ Thus if the argument difference of the two complex
coefficients is an acute angle, the state is entangled. Borrowing term from
optics, we can say that if the coherent addition constructively interferes,
the state is entangled.

\section{Conclusions}

We have introduced the ideas of coherent subtraction and coherent addition
to describe many kinds of bipartite non-Gaussian states and investigate
their inseparability. The non-Gaussian states produced from their Gaussian
kernels are quite generic. The results of the paper are of several folds.
For any coherent subtraction state produced form the second standard form
\cite{Duan} Gaussian state, we prove that it is always separable when the
Gaussian kernel is separable. For any symmetric coherent subtraction state
produced from the symmetric two-mode squeezed thermal state, the necessary
and sufficient separable condition is obtained and it is exactly the same as
that of its kernel state. A symmetric coherent subtraction state produced
from the symmetric first standard form Gaussian state (except two-mode
squeezed thermal state) is entangled when the Gaussian kernel is at the
boundary of separability and all the superposition coefficients have
positive real and/or imaginary parts. Thus a proper arranged coherent
subtraction produces entanglement from a separable state. For photon number
entangled state evolving in thermal noise and amplitude damping channel, we
give the analytical entanglement condition. As applications, we present the
separable conditions of two mode coherently correlated states with Poisson
coefficients and photonic Bell states evolving in thermal noise and
amplitude damping environment. Any two term coherent addition state produced
form boundary separable symmetric two-mode squeezed thermal state is
entangled if the two complex coefficients give rise to constructive
interference. Further works are desirable for other non-Gaussian states or
using entanglement criterion other than realignment criterion.

\section*{Acknowledgement}

This work is supported by the National Natural Science Foundation of China
(Grant No. 60972071), Zhejiang Province Natural Science Foundation (Grant
Nos. LQ12F01012,Y6110314)

\end{document}